# Excitation & Excavation of the Claws of the Southern Crab

Bruce Balick[1*], Ashley Swegel[1] and Adam Frank[2]
[1] Department of Astronomy, University of Washington, Seattle, WA 98195-1580, USA; balick@uw.edu, ashleyswegel@gmail.com
[2] Department of Physics and Astronomy, University of Rochester, Rochester, NY 14627, USA; afrank@pas.rochester.edu

## Abstract

We show that the Southern Crab (aka Hen2–104) presents an auspicious opportunity to study the form and speed of the invisible winds that excavate and shock the lobes of various types of bipolar nebulae associated with close and highly evolved binary stars. A deep three-color image overlay of Hen2–104 reveals that the ionization state of its lobe edges, or "claws", increases steadily from singly to doubly ionized values with increasing wall latitude. This "reverse" ionization pattern is unique among planetary nebulae (and similar objects) and incompatible with UV photoionization from a central source. We show that the most self-consistent explanation for the ionization pattern is shock ionization by a fast (~600 km s$^{-1}$) "tapered" stellar wind in which the speed and momentum flux of the wind increase with equatorial latitude. We present a hydrodynamic simulation that places the latitude-dependent form, the knotty walls, and the reverse ionization of the outer lobes of Hen2–104 into a unified context.

*Unified Astronomy Thesaurus concepts:* Planetary nebulae (1249); Symbiotic binary stars (1674); Stellar winds (1636); Shocks (2086)

## 1. Introduction

Post-AGB bipolar nebulae consist of one or more pairs of open or closed lobes that have been shaped by winds from their central stars and compact companions as they approach the final stages of their evolution. One subclass of bipolar nebulae, known as "symbiotic bipolars" (Whitelock & Munari 1992; hereafter "SymBNe"), is associated with dusty post-AGB stars. The nebular gas is ejected from its putative binary central star (hereafter "CS") consisting of a cool AGB star or Mira (color temperature ≈ 3000K), with a white-dwarf (hereafter "WD") companion. The binary CS is engulfed in a very dense and spatially unresolved cocoon of gas (electron density $n_e \gtrsim 10^6$ cm$^{-3}$) and dust from which a wide range of highly excited permitted and forbidden lines emerge. The surrounding spatially resolved nebula emits nebular emission lines in low-ionization lines, such as [S II] and [N II], and weak [O III]$\lambda$5007Å.

The nebular gas of SymBNe consists of the remnant mass of slow winds originally ejected from the cool AGB star (Höfner & Olofsson 2018, Höfner & Freytag 2022). Subsequently the nebula is reshaped by collimated fast winds (>100 km s$^{-1}$) from a putative dense accretion disk around the nearby companion WD (e.g., García-Segura et al 2018, Clyne et al. (2015; hereafter "C+15"), Santander-García et al. 2008, hereafter "S+08", Corradi & Schwarz 1993, hereafter "CS93"). The collimated and highly supersonic "fast winds" compress, shock, and accelerate the visible edges of the lobes (C+15; see also CS93, Corradi et al. 2001, hereafter "C+01", Hrivnak et al. 2008, Santander-García et al. 2004). If the fast winds are "light" [1] (Akashi & Soker 2008) then the speeds of the emergent winds within the seemingly hollow lobes[2] may be even larger than that seen in emission features within the lobe walls.

---

[1] A light wind is less dense than the external gas at the same radius and retarded by the inertial resistance of its environment. The shape of the lobe that it creates differs from the geometric form of the winds. In contrast, a "heavy" wind, such as a very dense jet, generally plows through the ambient medium without slowing down and the edges of its lobes reflect its form (Akashai & Soker 2008).

[2] The presence of fast stellar winds is directly inferred from the presence of wide photospheric absorption lines formed in the spatially unresolved wind acceleration zone just above the stellar surface where dust particles first form. Soon thereafter the winds encounter and shock a very dense circumstellar cocoon which modifies their speed and momentum. The geometry and the ultimate speeds of the diverging and optically invisible winds that reach and compress the lobe walls of SymBNe is unobserved.

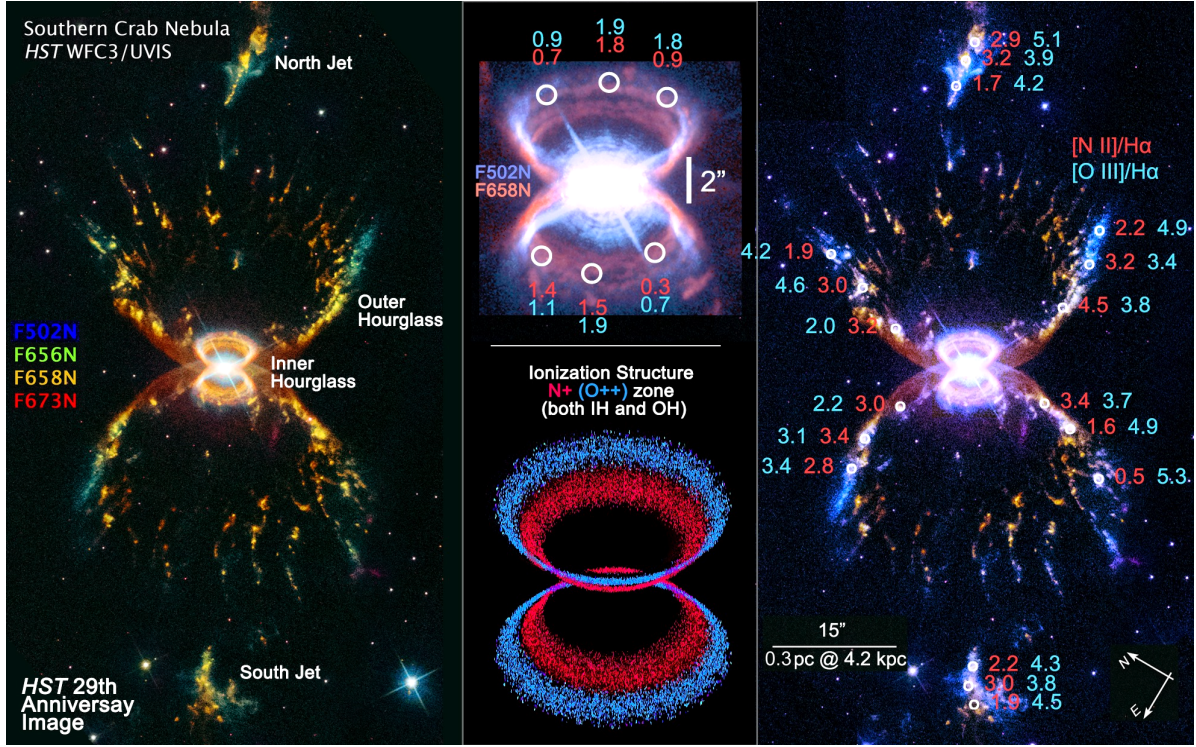

Figure 1. Left: HST image of Hen2–104 obtained by GO 15677 and published by NASA with the indicated color scheme as a press release[1]. Far right: high-contrast overlay of the outer hourglass in [N II] (red), H$\alpha$ (green), and [O III] (blue). Doppler measurements show that the upper lobe is approaching. Top center: a 3x expanded high contrast overlay of the inner hourglass in [N II] (red) and [O III] (blue). The emission-line ratios of [N II]/H$\alpha$ ([O III]/H$\alpha$) within the white 1″-wide circles (~3000 au) are shown in red (cyan). Bottom center: schematic of the large-scale ionization structure of hourglass structures (the upper lobe is tipped forward).

We distinguish SymBNe (Whitelock & Munari 1992) from similar looking bipolar planetary nebulae (hereafter "BPNe") whose CS's are found on H-R tracks well beyond the AGB stage. Examples of BPNe are NGCs 6302, 2440, 6445, 6537, and 2818 (J. Kastner, private communication). He$^{++}$ emission lines are common in BPNe, so their CS's are much hotter than those in SymBNe. C+15 showed that SymBNe are readily separable from BPNe by their IR colors and O$^{++}$ to H$^{+}$ emission line ratios.

This paper focusses on the SymBN Hen2–104. It is one of a family of "crab-shaped" SymBNe, also including M2–9, BI Cru, R Aqr, Hubble 5, Mz3, and possibly MyCn18, (e.g., CS93, C+15) each with a D-type (dusty) symbiotic binary central star. All are on the order of 20-35″ in angular radius and exhibit outflow Doppler speeds, ~100–200 km s$^{-1}$ (CS93, Bryce et al. 1997, O'Connor et al 2000). At typical distances of 2-4 kpc (GAIA EDR3, Brown et al. 2021), the radius of their lobes is ≈0.4 pc, and their expansion ages are a few millennia. Their total nebular masses are unmeasured but likely ≳0.1 M$_\odot$ if they are comprised mostly of material ejected from the evolved AGB or Mira CS of initial mass ≈1 M$_\odot$.

Figure 1 shows images of the lobes of Hen2–104 obtained using the Hubble Space Telescope ("HST") in 2019 to honor its 29$^{th}$ anniversary[3]. At first glance it is obvious that the visible nebula consists of two sets of lobes, inner and outer, each with identical shapes but different expansion ages (C+04, S+08, C+15). Even more unusual, the pattern of the colors along the claw legs changes systematically with radius, from orange to cyan in the Hubble public release image. This reflects a secular increase from predominantly singly to doubly ionization states of N and O. with latitude.

The color gradient shows that doubly ionized ions such as O$^{++}$ lie in zones lying *beyond* singly ionized species such as N$^{+}$. We show this more emphatically in the center and right panels of Figure 1. This "reverse" ionization pattern—the primary new result of this paper—is incompatible with photoionization models. It renders Hen2–104 unique among SymBNe and planetary nebulae ("PNe"). As we shall argue, reverse ionization is the outcome of "tapered" winds which impact the lobe edges with increasing speeds (from ~ 50 to ~150 km s$^{-1}$) with latitude. The pattern of shock excitation along its walls provides insight into the structure of the impinging stellar winds and, ultimately, the nature of the wind source.

---

[3] https://www.nasa.gov/image-feature/goddard/2019/hubble-celebrates-29th-anniversary-with-a-colorful-look-at-the-southern-crab-nebula

Another highly unusual feature of Hen2–104 is the outward-pointing tails associated with each of the many knots along the extraordinarily thin "claws" at mid latitudes of the outer hourglasses (Figure 1, see also C+15, Figure 9). The orientations of the radial tails also suggest that they are likely formed and shaped by invisible and fast radial winds from the core that sweep past them. In contrast, the tails of nearly all knots in PNe, such as the bullets in CRL618 and Ori-LMC, appear as contrails lying inside (towards the nucleus) of each their respective knots at their heads.

The data included in this paper, their calibration, our methodology of analysis, and results obtained directly from them are described in section 2. In section 3 and the Appendix we present and discuss the relationship between the large- and small-scale ionization structure and kinematics within the lobe walls of Hen2–104 and the form of the fast winds that might shape them, with emphasis on the better observed outer lobes. A 2-D numerical hydro simulation that reflects the large- and small-scale morphology and kinematics of Hen2–104 is presented. We draw general conclusions in section 4. For brevity we designate the inner and outer hourglass lobe pairs as the "IH" and "OH", respectively.

## 2. Nebular Structure and Kinematics

Our immediate goal is to use the multi-epoch, multi-filter HST images of Hen2–104 (1) to measure the ionization patterns of the line ratios in the lobe walls of the OH at mid latitudes where sharply defined knots are most prominent and (2) to extract the patterns of proper motions of these knots. This section is based mainly on fully calibrated long-exposure HST images extracted from the Hubble Legacy Archives using the Wide Field Camera 3 (WFC3) camera in 2019 and obtained in narrowband F502N, F656N, and F658N filters. The filters include the bright nebular lines of [O III], H$\alpha$, and [N II], respectively, near the centers of their passbands since the systematic speed of Hen2–104, $v_{\rm LSR}$, is modest: -85 km s$^{-1}$ (S+08). The exposure times are 3750s, 3100s, and 3100s, respectively. We also downloaded WFPC2 images of Hen2–104 in the F658N filter from 1999, 2003, and 2019 for our proper-motion studies.

All HST images downloaded for this study had been corrected for CCD bias, dark current, shutter shading, flat-field and charge transfer nonlinearities, geometric distortion, and cosmic rays. We aligned the images to the CS and transformed the pixel values from units of detected counts s$^{-1}$ pixel$^{-1}$ to absolute flux using standard HST filter throughput calibrations. Contamination by other lines and the nebular continuum are insignificant compared to the noise level in the images.

### *2.1 Spatial Features*

The key morphological features of Hen2–104 were first described by Schwarz et al. (1989) and Lutz et al. (1989, hereafter "L+89"). C+15 used the powerful 3-D modeling tool SHAPE model (Steffen et al. 2011) to develop a picture of the current spatio-kinematic relationships of the rings of the IH and the knots on the edges of the OH[4]. (We refer the reader to their very impressive results as presented in

---

[4] SHAPE assumes homologous expansion (which our proper-motion studies confirm), an inclination, and a user-supplied list of features and their projected locations in order to generate synthetic high-dispersion spectra that are readily compared to observed spectra (in this case, the Echelle observations in six slit orientations by C+01).

figures 5, 6, and 9.) S+08 and C+15 identified several additional features of Hen2–104 that will be important in section 3. These include:
1. the IH and the OH have the same shapes and symmetry axes but differ in size by a factor of ~5;
2. the lobes of the IH and OH appear to be open;
3. the edges of the OH are smooth in the equatorial zone (latitude $\phi \leq 17°$);
4. the knots and their tails of the OH lie at mid latitudes ($17° \leq \phi \leq 55°$);
5. the lobe walls become invisible at higher latitudes (roughly $\phi \gtrsim 55°$);
6. the knots within the walls of the OH are randomly distributed;
7. the rings in the walls of the IH lie at mid latitudes $25° < \phi < 39°$;
8. the entire nebula is inclined to the line of sight by ~59° (C+01, C+15);
9. the approaching NW (upper) lobe is tipped forward);
10. the nebula lies at a distance of 3.3±0.9 kpc (S+08);
11. the kinematic age of the OH is 4200±1200 y (S+08);
12. the kinematic age of the IH is 2.4 times smaller than that of the OH (C+15).

*2.2 Ionization Structure*

We measured the detected counts of [O III], H$\alpha$, and [N II] within 1″-diameter apertures (shown in Figure 1 as white circles) in the WFC3 images, converted them to fluxes using calibrations data supplied for each filter, and corrected the [O III] for reddening using the value of E(B–V) = 0.19 from Pereira (2005). The ratios of [N II]/H$\alpha$ and [O III]/H$\alpha$ are shown near the circles in red and cyan, respectively. Uncertainties in the line ratios are dominated by photon count noise.

Line ratios are useful diagnostics of local excitation conditions. As is obvious from the color changes in the claws seen in Figure 1, [N II]/H$\alpha$ systematically declines with latitude in both the IH and OH whereas [O III]/H$\alpha$ is steady or rises. Each mid-latitude knot shows an outward radial tail of higher ionization than the knot at its base. The schematic diagram shown in the lower center region of Figure 1 gives a simplistic but still useful description of the primary ionization zones, low (red) and high (cyan).

*2.3 The Jets*

A pair of knotty jets lie ±30″ from the center near the symmetry axis of Hen2–104 (Figure 1). The knots are prominent in all deep WFC3 images—especially in F658N—whereas more diffuse surrounding gas is best seen in F502N. The knots within the jets do not exhibit distinct tails; however, it is plausible that the tails are present but merge in space and/or overlap in projection.

*Supplementary Observations:* The slits used to observe the Echelle spectra published by C+01 and analyzed by C+15 were too short to cross the jets. Dr. Miguel García-Santander kindly supplied an unpublished Echelle spectrum obtained using the EMMI2 spectrograph on the ESO NTT telescope in 2008 (private communication). The jets clearly fall on the same average trend line of Doppler shift with radius as do the knots along the edges of the OH. Thus, the jets and the knots share a common expansion age (in disagreement with C+15 who surmised that the jets were four times younger than the OH).

*2.4 Patterns of Proper Motions*

The ensembles of knots and rings of Hen2–104 are brightest and sharpest in the F658N images in 1999 and 2019. Thus, we used this pair of images for measurements of proper motion studies. However, the rings in the IH are too poorly resolved to determine their proper motions. We applied the now-standard "magnification factor" analysis of Reed et al. 1999 to find the proper motions of the OH. In summary, we empirically found magnification a factor $M$ which, when applied to the 1999 image, gives a null residual when this image is subtracted from the 2019 image (modulo noise and field stars). Our result is a 20-year magnification factor of $M = 1.0005 \pm 0.0002$, or $2.5 \pm 1$ km s$^{-1}$ per arcsecond, at the distance of Hen2–104, in agreement with by estimates C+15. The corresponding expansion age, $T$, is $20$ y/$(M-1) = 4000 \pm 1000$ y[5].

We also divided the 2019 image by the magnified 1999 image. No signs of surface brightness changes or deviations from homologous expansion were detected over that time span.

*Supplementary Observations*: Proper motions were also measured by S+08; however, use of the newest WFC3 images make it possible to extend their four-year time-base to 20 years. Our results for $M$ agree well with theirs after adjusting for the different time spans, but with uncertainties half as large.

### 3. Deconstructing the Nebular Evolution

Our most significant observational finding is the increase of ionization with latitude from the CS of Hen2–104. Hereafter we use the term "reverse ionization" to describe it. *It is unprecedented among all classical PN of all morphological types that have been imaged by HST in the [NII], [O III] and Balmer emission lines.* Reverse ionization is also unique to Hen2–104 among other SymBNe—M2–9, Hubble 5, MyCn18, and Menzel 3—for which HST images have been obtained in the same emission lines.

We consider reverse ionization to be just a symptom whose underlying cause we explore in this section. As we shall see, Hen2–104 a potentially propitious case for unveiling the underlying nebular shaping agent in SymBNe and possibly other types of bipolar nebulae. We will also present a detailed hydro simulation in section 3.3 that matches the morphology of Hen2–104 and use it to form a coherent framework for all of the significant intrinsic features of Hen2–104 discussed earlier.

#### 3.1 Tapered Winds

"Tapered winds", are collimated stellar winds whose ram pressure and speed increase with equatorial latitude[6]. There is extant evidence of tapered flows in Hen2–104: S+08's analysis of their integral-field unit spectroscopy line led them to conclude: "Our spatio-kinematic fit clearly shows that the outflow speeds are slowest at low latitudes and highest along the polar direction". Specifically, they found that the nebular expansion speed in the polar direction, $v_{polar}$, and the equatorial plane, $v_{equator}$, are ~230 and ~12 km s$^{-1}$ for the outer lobes and ~90 and ~11 km s$^{-1}$, respectively, for the inner lobes. In addition, the speeds of relatively dense clumps immersed in light central winds from a central source will be a fraction of the local wind speed since the ram pressure of such winds rapidly declines with geometric dilution.

#### 3.2 Ionization Structure

One-dimensional models of photoionization (e.g., Osterbrock & Ferland 2006, Ferland et al. 2017) uniformly predict—and observations generally confirm—that high ionization zones lie interior to low ionization zones[7]. Models with irregular density distributions are unlikely to change the predicted global ionization structure of PNe since the transfer of UV radiation is primarily a 1-D process.

---

[5] Recently a value of $D = 4\ 2 \pm 0.08$ kpc to the CS of Hen2–104 was estimated by the GAIA (Brown et al.), in which case the S+08 method yields $T = 5500$ y.
[6] Tapered winds were originally introduced into studies of BPNe by Lee & Sahai (2003) in order to explain the formation of dense bullets of CRL618 and the closed candle-flame-like morphology and the increasing Doppler shifts of the highly compressed tails behind them.
[7] The primary reason why ionization levels drop with distance in model simulations is that the UV radiation density declines with distance faster than the particle density. Thus, the local recombination time per ion, $\approx 10^5$ y/$(Z^2 n_e)$, causes doubly charged ions to be depleted faster than singly ionized ions can be re-ionized. Here $n_e$ is the local electron density and $Z$ is the charge of the recombining ion.

Photoionization is generally assumed to be rule for the ionization and heating process in SymBNe, PNe, and other nebulae associated with post AGB stars[8]. This was justified by Dopita (1997) who showed that the global energy flux of the stellar UV radiation field is generally ~ 100 times larger than the input of mechanical energy across shocks. Moreover, the temperature of the hot star in the CS of Hen2–104, from 30,000˚K to 130,000˚K or more, (de Freitas & Costa 1996 hereafter dFC96", Kaler and Jacoby 1989), Contini & Formiggini, 2001; hereafter "CF01") is very typical of SymBNe and PNe.  (The luminosity of any unseen WD companion is unknown.)

Thus, some other global ionization process dominates in the claws[9]. The reverse ionization of Hen2–104 simply cannot be reconciled with the predictions of photoionization models.  In addition, the line ratios of Figure 1 lie at the periphery of hundreds of photoionized PNe in the [O III]/H$\alpha$ – [N II]/H$\alpha$ diagnostic diagrams by Frew & Parker (2010, Figure 5) and incongruent with photoionized H II regions in the "Baldwin-Phillips-Terlevich diagnostic plot" (Wang et al. 2017).

We explored whether the [O III]/H$\alpha$ and [N II]/H$\alpha$ ratios in Figure 1 might arise in shocks.  Raga et al. (2008, hereafter "R+08") published theoretical line ratios produced in shocked, fast-moving cloudlets, such as the FLIERs of NGC 6543, NGC 7027 and IC4634, whose emission-line ratios closely resemble those in the IH and OH.  The line ratios of Figure 1 are best matched by model A100 in which the shock speed, $v_s$, = 100 km s$^{-1}$ (and the adopted chemical abundances are solar).  Even so, typical measured [O III]/H$\alpha$ ([N II]/H$\alpha$) ratios and those in the FLIERS are discrepant from those in Table 1 of R+08 by factors ~5x (~2x).  On the other hand, the discrepancy of the line ratios predicted by R+08 for young, shocked cloudlets and the those observed in Hen2–104 are vastly smaller than the typical discrepancies same ratios of hundreds of photoionized HII regions and PNe in the diagnostic plots cited above.

This and the reverse ionization of the claws (and the IH) leaves little doubt that that emission from the claws comes from shock excitation.  But the evidence isn't quite fully decisive.  Certainly, uncertainties in our measurements are relevant.  Moreover, predictions of shocked line ratios simply depend on many parameters (cf. Hartigan et al. 1987), most of which aren't well known for Hen2–104.

*3.3 Physical Structure and Wind Shaping*

Fast collimated winds are a standard way to account for the formation of bipolar lobes, shocks, and the formation of rings and knots along their outer walls (e.g., Lee & Sahai, 2003, Chita et al. 2008).  The ram pressure of winds on the outer walls of the lobes creates a thin shell that can be rendered unstable to wind-induced structural perturbations and R-T instabilities (McLeod and Whitworth 2013).

Moreover, tapered winds in this model readily explain the curvature of lobe walls[10] and the systematic increase of ionization (i.e., shock speed) along the walls.  Specifically, the winds strike and shock the swept-up and compressed walls at maximum speed in the polar zone and steadily more slowly at lower latitudes. In addition, the effective shock speeds depend on the local of the obliquity of the wind impact angle, $\varphi$ (the effective shock speed scales with sin $\varphi$). The increasing wind impact speed and angle with latitude accounts for the observed variations of [O III]/H$\alpha$ and [N II]/H$\alpha$ found along the claws of the IH and OH.

---

[8] Evidence of shocks within primarily photoionized PNe is emerging.  P. Moraga (private communication) has recently uncovered evidence of shocks at the interfaces of some major morphological structures in the iconic PN NGC 7027.  Kastner et al (2022) and Smith et al. (2005) found lines of [Fe II]1.64μm, a shock tracer, along the lobe walls of NGC 6302 and M2–9, respectively.  See also Lago et al. (2019).

[9] Early warnings of problems of photoionization of Hen2–104 were noted by L+89 who wrote "The 'crab legs' are far enough away from the hot source that their spectrum comes from a combination of photoionization and shock."  CF01 reached the same conclusion based on a more thorough analysis.  Similarly, in their unsuccessful effort to estimate chemical abundances based on ionization models, dFC96 wrote "[the deviations of the derived predictions] could be an indication that some ions may be produced by shocks…".  However, it is now clear from the iron lines that the spectra considered by the authors were badly contaminated by shock-like emission lines arising in the very dense nucleus or accretion disk (C+01), not in the IH or OH.  So these early shortcomings can be dismissed.

[10] In the absence of tapering or weak magnetic fields the lobes would look like projections of cones with straight radial walls and a crust of swept-up gas along their leading edges.

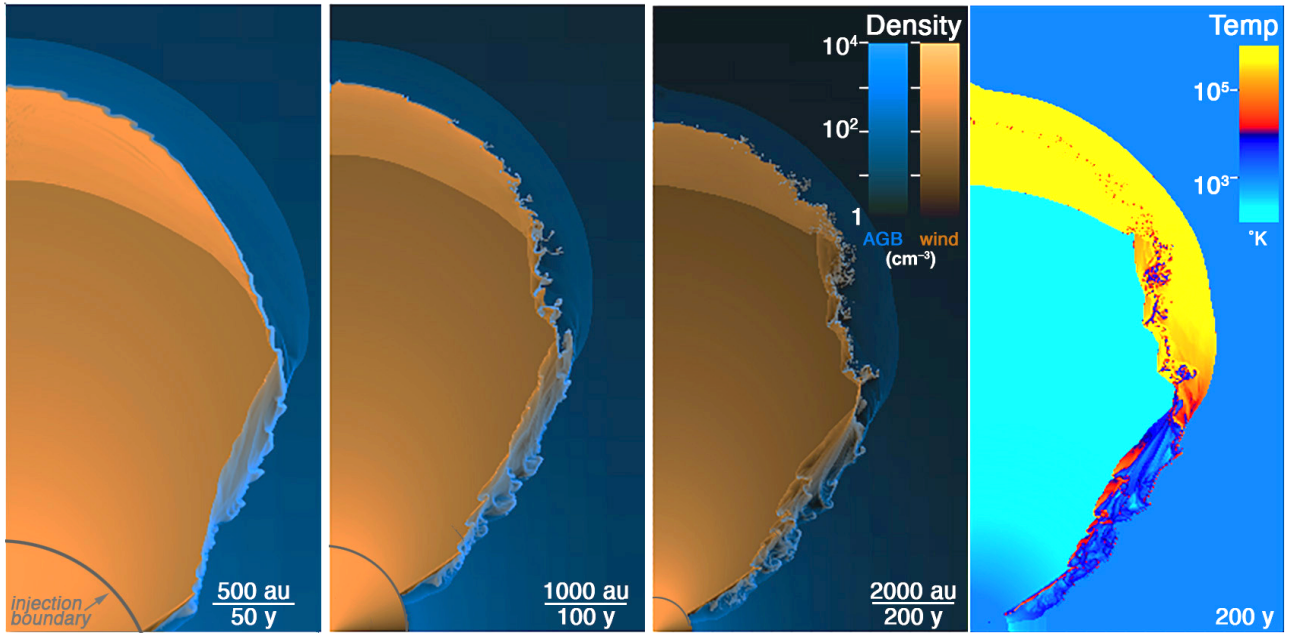

Figure 2. The three left frames show the density distribution of the hydrodynamic simulation at 50, 100, and 200 y. The tapered wind is injected from a round surface 200 au from the origin with an axial speed of 600 km s$^{-1}$. The blue (orange) colors show the densities of the ambient gas (injected fast tapered wind). The right frame shows the temperature at 200 y. Thermal pressure dominates in the (optically invisible) yellow-orange regions and ram pressure dominates elsewhere within the lobe. (Note that the dimensions of the first three panels scale linearly with time.)

While C+15's SHAPE model provides a snapshot of the current geometry of Hen2–104, hydro simulations are a means to develop insight its earlier evolution. Accordingly, we ran 2-D numerical hydro simulations with two goals in mind; to recreate the large-scale structure of Hen2–104 and to demonstrate how knots are likely to develop at mid-range latitudes in the thin walls of the lobes.

We used the program "AstroBear 2.0" developed at the University of Rochester (Cunningham et al. 2009, Balick et al. 2013). We injected a steady, cold, radial fast wind with a Gaussian-tapered opening angle at 1/$e$ width = 50˚ and an axial speed of 600 km s$^{-1}$ into a cold, static ambient AGB wind with an inverse-square density profile. The wind is light: the initial ratio of fast-to-AGB wind densities is ½ at the injection surface. The smallest grid resolution, $\delta r$, = 2 au. Details of the model are found in the Appendix.

We ran about 100 simulations with different initial conditions and terminated the model as the lobes reached the edges of our 64 × 128 kau (0.3 × 0.6 pc) computational domain. The best of these, shown in Figure 2, nicely matches the global shape and expansion pattern of the HST images and our proper-motion studies after 200 y, after which the lobe shape becomes fixed. Running the models for a longer time seemed pointless since a 3-D domain (with three degrees of freedom) is needed to follow the realistic development and evolution of thin knots, clumps, and rings within the claws. Thus, the present simulations are simply illustrative.

Note that that the lobe walls quickly separate into three latitude zones. These correspond approximately to the three major latitude zones of Hen2–104 found by C+15 and described in section 2: equatorial (relatively smooth lobe walls), mid-latitude (knotty walls), and polar (no walls). In the equatorial zone the speed that characterizes the shock in the equatorial zone is relatively small owing to the taper of the wind and its impact on the inner wall edges is highly oblique. We can expect that the shock speed to be lowest and the [O III]/[N II] line ratio weakest compared to zones at higher latitudes. Thus, the lower inner lobe walls are wide and relatively cool with minor K-H instabilities but no fragmentation.

By comparison, the effective speed of the tapered flow at mid latitudes is greater and the impact is less oblique. Hence the [O III]/[N II] line ratio is enhanced. The walls of the lobe rapidly fragment into knots consisting of displaced AGB gas. Their radial speeds (i.e., proper motions) increase monotonically with latitude up to ~200 km s$^{-1}$. Only the cooler knots (blue in the rightmost panel) are optically visible.

The reverse shock upstream from the locus of knots is pushed inward by its thermal pressure, so the knots are increasingly isolated from the ram pressure of the wind and stabilized by the surrounding thermal pressure. The knots attain their relative positions within the lobe walls within $t \sim 200$y, thereby accounting for the homologous expansion pattern of their observed proper motions. 3-D models with an extra degree of freedom may allow ablation tails to form (as seen in simulations of widely separated knots immersed in high-speed light winds by Steffen et al. 2004).

The polar zone may be optically invisible, but it isn't empty. The average wind speed in this zone is >500 km s$^{-1}$, so the original walls of the lobe quickly "crush" and dissipate into sparse, hot (>10$^7$K) shell of gas containing some embedded warm vestigial "droplets" of cooler gas at the contact discontinuity. (Crushing is described in the Appendix.) Elsewhere the thermally expanding AGB winds encounter little inertial resistance, so the leading edge of the lobe rapidly expands into the sparse gas downstream. Therefore, the nebular gas within the polar zone resembles a cone-shaped segment of a hot wind-blown bubble with smooth leading and reverse shocks on their forward and lee edges (e.g., Toalá & Arthur 2014). Optically, the lobe appears open. Soft X-rays could be very faintly visible in its denser regions.

Beyond the lobes, the decreasing density and increasingly feeble inertial resistance of the downstream AGB winds cannot inhibit the motions of the dense knots after about 100 y. Thus, we expect the relative locations and the radial speeds of the structures seen at 200 y to remain the same indefinitely. The exception is the region of very hot and sparse gas in the polar zone which emulates a conical slice of a loosely constrained hot bubble.

## 4. Discussion and Conclusions

Our principal conclusion in this paper is that the nebular gas in the walls of Hen2–104 is ionized and heated by fast stellar winds, not stellar UV photons like virtually all similar nebulae associated with highly evolved stars. This conclusion is readily consistent with multiple lines of evidence, including the general bipolar shape of the OH, it's very thin knotty walls, and the "reverse" spatial ionization pattern of the lobe walls from low to mid latitudes. Many of these attributes also apply to the IH, suggesting that the IH is following a similar path of evolution as the OH, albeit at ~40% of its age.

A 2-D hydro model based on fast tapered stellar winds flowing into initially denser and slower AGB winds nicely matches the global physical structure of Hen2–104. It's latitude-dependent flow speed accounts for the odd changes of the ionization, or 'reverse ionization", seen along its lobe walls at mid latitudes. Therefore, the initial conditions that we adopted for the fast tapered winds in this simulation (see the Appendix) provide insight into the geometric form and speed of the winds emerging from its nucleus. Tapered winds also explain the seemingly open geometry of the lobes at high latitudes as well as the smooth structure of the walls at low latitudes. The hydro model does not explain the knot tails. 3-D simulations, possibly with lightly modified initial conditions, are probably required to this end.

We are unable to say how the continuous and evenly spaced lobe rings in the IH originated. It is possible that they are the result of a companion in a highly elliptical orbit that plunges into and ejects portions of the loosely bound Mira star's atmosphere every few centuries (Soker & Kashi 2012).

Observations of the lobe edges of the IH and OH of Hen2–104 directly support lobe formation by tapered winds (S+08), of which there have long been strong hints in other examples of other bipolar nebulae (e.g., Lee & Sahai 2003, Balick et al. 2019). Tapering places strong constraints on stellar wind formation mechanisms. A viable source of such winds is axial outflows from accretion disks surrounding the compact fed by a star with a loosely bound atmosphere, such as AGB or Mira. (It is tempting to speculate that thin jets from a precessing accretion disk inscribe closely spaced spiral rings in the lobe walls of the IH.) Moreover, the collimated winds must be long-lived owing to the persistence of shocked regions from which emission lines are still clearly visible after several millennia.

Hen2–104 is the clearest case of a shock-ionized SymBN. But it may not be the only one. A study of the SymBN BI Cru whose CS is a relatively hot Mira CS (26500˚K) by Contini et al. (2009, "C+09") concluded that it is predominantly shock ionized. BI Cru shares many of the same morphological and kinematic similarities with Hen2–104 (CS93) and other SymBN and BPNe. One the other hand, their conclusion is based on spectral line strengths that are dominated by shocked circumstellar stellar gas.

Finally, W-R nebulae are a useful point of comparison. Gruendl et al. (2000) obtained HST F502N and F656N images of eight W-R nebulae. Six of them show very thin rims of enhanced [O III]/H$\alpha$ ratios at their outer edges with lower ionization structures in their interiors. They suggested that the [O III]/H$\alpha$ enhancement of the rims arises in post-shock recombination zones driven into the pre-existing ISM by thermal pressure. Indeed, Figures 8 – 10 in Toalá and Arthur (2011) show the production of knots with radial tails formed by thermal pressure along the shocked rims of W-R nebulae.

We are especially indebted to Dr. Miguel Santander-García for providing the data shown in Figure 2 which he obtained using ESO's NTT telescope with the EMMI2 spectrograph. Baowei Liu played an essential role in running the numerical simulations. Joel Kastner supplied lengthy AstroPy routines that we modified to derive our proper motion results. Jesus Toala kindly provided valuable insights into W-R nebulae. We thank the referee for a meticulous, constructive, and thorough reading of the manuscript.

This research is partially based on observations made with the NASA/ESA Hubble Space Telescope obtained in program GO15677 and downloaded from the MAST data archive, both which are operated by the Association of Universities for Research in Astronomy, Inc., under NASA contract NAS 5–26555. HLA is a collaboration between the Space Telescope Science Institute (STScI/NASA), the Space Telescope European Coordinating Facility (ST-ECF/ESA) and the Canadian Astronomy Data Centre (CADC/NRC/CSA). This work has also made use of data from the European Space Agency (ESA) mission Gaia (https://www.cosmos.esa.int/gaia), processed by the Gaia Data Processing and Analysis Consortium (DPAC, https://www.cosmos.esa.int/web/gaia/dpac/consortium). Funding for DPAC has been provided by national institutions participating in the Gaia Multilateral Agreement. AF acknowledges support grants DE-SC0001063, DE-SC0020432, DE-SC0020434 from the Department of Energy and grant AST-1813298 from the National Science Foundation.

*Facilities:* HST (WFPC2 and WFC3)

### ORCID IDs

Bruce Balick https:/orcid.org/0000-0002-3139-3201
Adam Frank https:/orcid.org/0000-0002-4948-7820

## Appendix: Numerical Simulations

Figure 2 shows the outcome of injecting a fast and steady, "light" tapered wind[1] into a stationary inverse-square density distribution, $n_{AGB}$, that represents the slow, ambient AGB wind. The density, $n_w$, and speed, $v_w$, of the radially diverging fast wind are modulated in latitude by a Gaussian with a 1/e-width of 50°. Here $n_w(r_o)$ =1000 cm$^{-3}$, $n_{AGB}(r_o)$ = 2000 cm$^{-3}$, $v_w(r_o)$ = 600 km s$^{-1}$, where $r_o$ = 200 au is the radius of the wind injection boundary. The distributions of gas originating in the fast wind (orange) and the much slower and denser AGB gas (blue) are shown at 50, 100, and 200 y. The outflow speeds of the radiating knots are ≈50% of the speeds of the fast winds at the same latitude.

Thin-shell instabilities are initiated in the lobe walls at mid latitudes by the ram pressure of the stellar winds by $t$ = 50 y. Knot formation is well underway within $t \sim 100$ y and ends by ~200 y when a sheath of hot gas (≈10$^7$°K) forms around the knots. The sheath buffers the knots from the ram pressure of the fast stellar winds while the thermal pressure of the adjacent gas confines them. At the same time, a slow leading shock ($v_s \sim 30$ km s$^{-1}$) forms at the interface of the expanding lobes with the undisturbed ambient gas downstream. (The leading shock may be observable in excited lines of H$_2$ and other molecular tracers.) The hydrodynamic behaviors in other latitude zones are presented in section 3.

The model of Figure 2 is simply illustrative. Obviously, there may exist a family of similar hydro models that fit the shape and dynamics of Hen2–104 as well as or better than that of Figure 2. We have not explored the outcomes for different power-law forms of the ambient AGB gas downstream or for different injection density contrasts $n_w(r_o)/n_{AGB}(r_o)$ Also, we have not run simulations with faster winds $v_w(r_o) > 600$ km s$^{-1}$.

Moreover, we have not run models designed to fit the growth rate of the IH and the formation of its rings since it is even more poorly constrained by observations than is the OH. Also, our 2-D model fails to develop the radial tails downstream of the knots. See also roughly similar models by Steffen & López (2004) in which straight tails form between their widely separated knots.

"Crushing" can be an important factor in the evolution of the knots and rings. Crushing is induced as shocks traverse clumps or filaments of cold gas on a sound-crossing time scale (Cunningham et al. 2009). The crushing time, $t_{crush}$, is $(2R_0/v_s) \cdot (n_r/n_w)^{1/2}$, where $R_0$ is the knot or ring thickness, $v_s$ is the speed of the shock, $n_r$ is the internal density of the ring, and $n_w$ is the density of the wind). If we apply this to the rings of the IH and adopt $n_r \sim 600$ cm$^{-3}$ (S+08), $R_0 \sim 0.''9$ (S+08) = 5.7×10$^{11}$ km, $v_s \sim v_w \sim 600$ km s$^{-1}$, and $t_{crush}$ = 2300 y (the age of the IH, C+15), then we derive a reasonable value for $n_w \lesssim 0.6$ cm$^{-3}$, albeit with large uncertainties in all parameters. Taken at face value, the integrity of the rings is not likely to persist much longer.

C+15 noted an abrupt change in the slope of the lobes at the base of the mid-latitude zone of the OH. This "elbow" has an obvious counterpart in the final frame of the model in Figure 2. Perspicaciously, they speculated that the rings of the IH of Hen2–104 might "break into smaller scale knots due to [pressure instabilities]" that may eventually resemble the knotty form of the OH. An F658N image of the outer lobe rings of MyCn18 (Sahai et al. 1999), another crab shaped SymBN with rings in its lobes, shows that they may show insipient signs of crushing into knots.

Limitations: (1) The diameters of the knots of Hen2–104 are ≈1″ (3 au). Creating such small knots with tails requires simulations in three model dimensions on a grid with smaller computational cells. This was beyond our means. (2) a shortcoming is that the model predicts an expansion coefficient of 4 km s$^{-1}$ after 4200±1200 y compared to an observed value of 2.5 km/s per arcsec (C+15). A kinematic age of 3000 y and/or a slower wind speed (450 – 500 km s$^{-1}$) resolve the disparity.